\documentclass[iop,tighten]{emulateapj}

\usepackage{subfigure}
\def\simgt{\lower 2pt \hbox{$\, \buildrel {\scriptstyle >}\over{\scriptstyle \sim}\,$}}
\def\simlt{\lower 2pt \hbox{$\, \buildrel {\scriptstyle <}\over{\scriptstyle \sim}\,$}}

\begin{document}

\title{X-ray constraints on the local supermassive black hole occupation fraction}

\author{Brendan~P.~Miller,$^{1,2}$ Elena~Gallo,$^{1}$
  Jenny~E.~Greene,$^{3}$ Brandon~C.~Kelly,$^{4}$ Tommaso~Treu,$^{4}$
  Jong-Hak~Woo$^{5}$, \& Vivienne Baldassare$^{1}$}

\footnotetext[1]{Department of Astronomy, University of Michigan, Ann
  Arbor, MI 48109, USA}

\footnotetext[2]{Department of Physics and Astronomy, Macalester
  College, Saint Paul, MN 55105, USA}

\footnotetext[3]{Department of Astrophysics, Princeton University,
  Princeton, NJ 08544, USA}

\footnotetext[4]{Physics Department, University of California, Santa
  Barbara, CA 93106, USA}

\footnotetext[5]{Astronomy Program, Department of Physics and
  Astronomy, Seoul National University, Seoul, Korea}

\begin{abstract}

Distinct seed formation mechanisms are imprinted upon the fraction of
dwarf galaxies currently containing a central supermassive black
hole. Seeding by Pop III remnants is expected to produce a
higher occupation fraction than is generated with direct gas collapse
precursors. {\it Chandra\/} observations of nearby early-type galaxies
can directly detect even low-level supermassive black hole activity,
and the active fraction immediately provides a firm lower limit to the
occupation fraction. Here, we use the volume-limited AMUSE surveys of
$\sim$200 optically-selected early-type galaxies to characterize
simultaneously, for the first time, the occupation fraction and the
scaling of $L_{\rm X}$ with $M_{\rm star}$, accounting for intrinsic
scatter, measurement uncertainties, and \hbox{X-ray} limits. For
early-type galaxies with $M_{\rm star}<10^{10} M_{\odot}$, we obtain a
lower limit to the occupation fraction of $>$20\% (at 95\%
confidence), but full occupation cannot be excluded. The preferred
dependence of $\log{L_{\rm X}}$ upon $\log{M_{\rm star}}$ has a slope
of $\sim$0.7--0.8, consistent with the ``downsizing'' trend previously
identified from the AMUSE dataset, and a uniform Eddington efficiency
is disfavored at $\sim$2$\sigma$. We provide guidelines for the future
precision with which these parameters may be refined with larger or
more sensitive samples.

\end{abstract}


\section{Introduction}

Observations of high-redshift quasars indicate that supermassive black
holes (SMBHs\footnote{We use the term ``supermassive'' to indicate
  masses of $M_{\rm BH}\simgt3\times10^{5} M_{\odot}$ for central
  black holes, as in Greene (2012).}) are already present in the early
universe (e.g., Vestergaard \& Osmer 2009; Willott et al.~2010;
Mortlock et al.~2011). SMBHs with $M_{\rm BH}\simgt10^{9} M_{\odot}$
by $z\simgt6$ are extremely challenging to grow from Population III
remnants (``light'' seeds of $\sim100 M_{\odot}$; e.g., Whalen \&
Fryer 2012; Madau et al.~2014; Taylor \& Kobayashi 2014), but can
derive from direct gas collapse precursors (``heavy'' seeds of
$\sim10^{5} M_{\odot}$; e.g., Begelman 2010; Johnson et al.~2013;
Ferrara et al.~2014). However, the unresolved cosmic \hbox{X-ray}
background implies SMBHs are not common (or else are generally
quasi-quiescent) in high-redshift galaxies (Salvaterra et al.~2012), a
possibility also suggested by stringent constraints on average nuclear
X-ray luminosities obtained from stacking deep field {\it Chandra\/}
observations (Triester et al.~2013). For typical expected subsequent
black hole growth (Shankar et al.~2013), and in line with the SMBH
mass function derived from broad-line quasars (Natarajan \& Volonteri
2012), these results may be more consistent with sparse heavy seeding
than with slow initial growth of omnipresent light seeds. Despite
significant and ongoing theoretical and observational advances, the
particular seed mechanism predominantly responsible for SMBH formation
is not yet conclusively established (see reviews by Volonteri 2012;
Volonteri \& Bellovary 2012; Natarajan 2014; and references therein).

The evolution of SMBHs appears to be entwined with that of their host
galaxies. This is suggested by the $M_{\rm BH}-\sigma$ relation
linking the central black hole mass to the bulge stellar velocity
dispersion, which holds for both quiescent spheroids (G{\"u}ltekin et
al.~2009; McConnell \& Ma~2013) and active galactic nuclei (Woo et
al.~2010, 2013) and may be redshift-dependent (Treu et al.~2007; Lapi
et al.~2014). SMBH feedback provides one plausible linking mechanism
(Sun et al.~2013), as predicted by semi-empirical modeling (Croton et
al.~2006; Shankar et al.~2013) and in a few cases now directly
measured (e.g., Feruglio et al.~2010; Cano-D{\'{\i}}az et al.~2012;
Liu et al.~2013). Mergers and intermittently efficient accretion in
larger SMBHs spur growth and remove observational signatures of their
birth, but smaller SMBHs have more subdued histories and undergo
mostly secular evolution (Jiang et al.~2011). Consequently, both the
mass distribution and the very rate of occurance of SMBHs in
lower-mass galaxies contain archaeological information on the initial
seed formation mechanism.

A robust conclusion from semi-analytical modeling is that smaller
galaxies are more likely to contain SMBHs when Pop~III remnants,
rather than direct gas collapse, provide the
dominant\footnote{Intermediate mass seeds, for example from nuclear
  star cluster collapse (Davies et al.~2011; Lupi et al.~2014), are a
  third possibility.} seeding mode (Volonteri \& Natarajan 2009;
Volonteri 2010; van Wassenhove et al.~2010). This is because cold
low-metallicity gas is only able to collapse to a central massive
object in halos with low spin parameter, otherwise disk fragmentation
leads to star formation (van Wassenhove et al.~2010). The fraction of
halos forming such heavy seeds should exceed 0.001 to produce SMBHs at
$z=6-7$ (Petri et al.~2012). Using a First Billion Years cosmological
hydrodynamical simulation, Agarwal et al.~(2014) identify several
pristine\footnote{Enriched gas cannot directly collapse to produce a
  massive seed (e.g., Ferrara et al.~2013).} atomic-cooling haloes
that could host direct-collapse massive seeds, and note that these
haloes are universally close to protogalaxies and exposed to a high
flux of Lyman-Werner radiation (as also found by, e.g., Latif et
al.~2013a, b; Dijkstra et al.~2014). Measurement of the occupation
fraction (i.e., the percentage of galaxies hosting SMBHs) in nearby
galaxies, particularly at low stellar masses $M_{\rm star}<10^{9-10}
M_{\odot}$, is an effective {\it observational\/} discriminator
between light versus heavy seeds (Greene 2012).

The limited $\simlt$10$^{8}$ yr lifetime of luminous quasars suggests
(Soltan 1982; Yu \& Tremaine 2002), consistent with observations, that
the most massive ``inactive'' galaxies invariably host SMBHs now
accreting/radiating only at $\simlt10^{-5}$ Eddington, but the
occupation fraction in lower mass galaxies remains uncertain. Clearly
some low-mass galaxies do possess SMBHs\footnote{The masses of central
  black holes in dwarf galaxies are difficult to measure precisely,
  but the following examples are likely near or above our adopted
  definitional threshold for a SMBH.}, even active ones. For example,
the dwarf galaxy Henize 2-10 hosts an accreting SMBH as revealed by
\hbox{X-ray} and radio emission (Reines et al.~2011), and features
central blue clumps of star-formation within a red early-type system
(Nguyen et al.~2014). Mrk 709 is an interacting pair of dwarfs, the
Southern of which has a central \hbox{X-ray} and radio source
indicating the presence of a SMBH (Reines et al.~2014). Within the
{\it Chandra\/} Deep Field South Survey, Schramm et al.~(2013)
identify three galaxies with $M_{\star}<3\times10^{9} M_{\odot}$ that
have \hbox{X-ray} emitting SMBHs. Yuan et al.~(2014) describe four
dwarf Seyferts with $M_{\rm BH}\simlt10^{6} M_{\odot}$, two of which
are detected in X-rays with $L_{\rm X}\sim10^{41}$~erg~s$^{-1}$. A
sample of 151 dwarf galaxies with candidate SMBHs as identified from
optical emission line ratios and/or broad H$\alpha$ emission is
presented by Reines et al.~(2013; see also references therein). The
ultra-compact dwarf galaxy M60-UCD1 is indicated by a central velocity
dispersion peak to have a SMBH with $M_{\rm BH}=2.1\times10^{7}
M_{\odot}$, but here the large black hole mass fraction suggests
substantial stellar mass has been stripped from the galaxy (Seth et
al.~2014). For each example of a low-mass galaxy that has
observational evidence for a central SMBH, there are 10--100 similar
galaxies for which the presence or absence of a black hole is
currently impossible to measure. However, dynamical mass constraints
are quite stringent for some Local Group objects (the spiral M33:
Gebhardt et al.~2001; Merritt et al.~2001; the dwarf elliptical
NGC~205: Valluri et al.~2005), which effectively rules out a 100\%
SMBH occupation fraction.

High spatial resolution \hbox{X-ray} observations can efficiently
identify very low-level SMBH activity (Soria et al.~2006; Pellegrini
2010) without contamination from the stellar emission that dilutes
optical searches. Nuclear \hbox{X-ray} emission directly measures
high-energy accretion-linked radiative output and additionally serves
as a plausible proxy for mechanical feedback (Allen et al.~2006;
Balmaverde et al.~2008). \hbox{X-ray} studies of low-level SMBH
activity are best conducted on galaxies with low star formation rates
to eliminate potential contamination from high-mass \hbox{X-ray}
binaries. For statistical purposes the sample must span a wide range
in $M_{\rm star}$ and be unbiased with respect to optical or
\hbox{X-ray} nuclear properties. These criteria are satisfied by the
AMUSE\footnote{AGN Multiwavelength Survey of Early-Type
  Galaxies}-Virgo (Gallo et al.~2008, 2010; G08, G10 hearafter) and
AMUSE-Field (Miller et al.~2012a, 2012b; M12a, M12b hereafter)
surveys, which are Large {\it Chandra\/} Programs that together
targeted 203 optically-selected early-type galaxies at $d<30$~Mpc, and
now include {\it HST\/}, {\it Spitzer\/}, and {\it VLA\//JVLA}
coverage. Almost all of these galaxies have $L_{\rm
  X}<10^{41}$~erg~s$^{-1}$ and $L_{\rm X}/L_{\rm Edd}<10^{-5}$, below
limits commonly used to distinguish active galactic nuclei (AGN) from
``inactive'' galaxies.

In this work we use the AMUSE dataset to obtain the first simultaneous
constraints upon the scaling of nuclear activity with host galaxy
stellar mass and the local supermassive black hole occupation
fraction, and derive guidelines for the precision that may be achieved
with a larger sample or a next-generation \hbox{X-ray} telescope.

\section{Disentangling occupation and downsizing}

The AMUSE nuclear detection fractions constitute, after correcting for
potential minor low-mass \hbox{X-ray} binary contamination (G08; G10;
M12a), strict lower limits on the occupation fraction. Taking as given
that all higher-mass early-type galaxies host SMBHs, the efficiency
with which their nuclear \hbox{X-ray} sources are found suggests a
correction factor to apply to the lower-mass galaxies. After assuming
a uniform distribution of Eddington-scaled luminosity, the occupation
fraction can be calculated in a straightforward fashion by imposing a
limiting Eddington sensitivity threshold. This approach tentatively
favors heavy seeds (Greene 2012).

However, the assumption of a mass-independent Eddington ratio
distribution is disfavored by the data. Both the Virgo and Field
galaxies display an apparent ``downsizing'' trend (with a consistent
slope) toward relatively greater Eddington-scaled \hbox{X-ray}
luminosity in lower-mass galaxies or for inferred lower-mass SMBHs
(G10; M12a; M12b). While this downsizing tendency is qualitatively
similar to the effect found at higher masses and accretion rates for
quasars, the physical explanation may be completely distinct, given
the very low accretion rates and radiative efficiencies that
characterize the AMUSE sample (including M~87, which has a mass
accretion rate directly constrained by the rotation measure to be two
orders of magnitude below Bondi; Kuo et al.~2014). Thus, we cannot
make direct comparisons with recent results questioning downsizing in
moderately luminous AGN with $42<\log{L_{\rm X}}<44$ (Aird et
al.~2012). In general, the presence of both downsizing and occupation
fraction complicates estimates of either parameter alone. For example,
a downsizing-enhanced detectability of SMBHs down the mass scale could
bias high an estimate of the occupation fraction that presumes a
uniform Eddington fraction. The slope of the dependence of $L_{\rm X}$
upon $M_{\rm star}$ is primarily sensitive to the higher mass
galaxies, most of which are \hbox{X-ray} detected, but could
potentially be influenced by partial occupation in lower mass
galaxies. To date occupation fraction and downsizing have not been
simultaneously constrained.

To investigate the occupation fraction of SMBHs and simultaneously
their Eddington rates across the mass scale, we consider the
measurable distribution of \hbox{X-ray} luminosities as a function of
host galaxy stellar mass. Motivated by prior studies we take
$\log{L_{\rm X}}$ to be a linear function of $\log{M_{\rm star}}$ but
allow for significant intrinsic scatter; see Figure~1a. It is assumed
that the degree of intrinsic scatter remains constant across the mass
scale and we note as a caveat that this is observationally
uncertain. The $L_{\rm X}$--$M_{\rm star}$ correlation is
observationally truncated by the sensitivity limit of the AMUSE
surveys, which is $\log{L_{\rm X}}\simeq38.3$~erg~s$^{-1}$. A
decreasing occupation fraction toward lower $M_{\rm star}$ would
result in a portion of galaxies not following the $L_{\rm X}$--$M_{\rm
  star}$ correlation but presenting instead as non-detections, since
they would lack an SMBH to generate \hbox{X-ray} emission.

We consider occupation fractions bounded by $f_{\rm occ}\simeq0$ for
$M_{\rm star}<10^{7} M_{\odot}$ and $f_{\rm occ}\simeq1$ for $M_{\rm
  star}>10^{10} M_{\odot}$. The probability of hosting an SMBH is taken
to be
\begin{equation}
0.5+0.5{\times}\tanh{(2.5^{|8.9-\log{M_{\rm
        star,0}}|}\log{\frac{M_{\rm star}}{M_{\rm star,0}}})}
\end{equation}
This simple functional form was selected because it is smooth and
spans a wide range of plausible possibilities, and in particular
includes both the light ``stellar death'' and the heavy ``direct
collapse'' competing seed formation models (from van Wassenhove et
al.~2010; Volonteri 2010) as illustrated in Greene~(2012; see their
Figure~2). We show this parameterization in Figure~1 for
$7.5<\log{M_{\rm star,0}}<10.2$, which correspond to occupation
fractions between 98\% and 2\% for galaxies with $M_{\rm star}<10^{10}
M_{\odot}$. Note that here and throughout occupation fractions are
derived from a given $M_{\rm star,0}$ value by applying the
probabilities in Figure~1b to the AMUSE $M_{\rm star}$ distribution in
Figure~1c, and that by construction even models with a low occupation
fraction (always quoted, we emphasize, for galaxies with $M_{\rm
  star}<10^{10} M_{\odot}$) produce nearly 100\% SMBH occupation in
high-mass galaxies.

The relationships we assume throughout between $M_{\rm star}$, $L_{\rm
  X}$, and the SMBH occupation fraction are illustrated in
Figure~1. The simulated sample of 10000 galaxies (the colored points
in the top panel) has $M_{\rm star}$ drawn from an unevenly-weighted
sum of four Gaussians constructed to empirically match the mass
distribution of the AMUSE surveys.\footnote{The central $\log{M_{\rm
      star}}$ values, standard deviations, and fractional weights for
  the four Gaussians are (7.9, 9.2, 10.3, 11.0), (0.2, 0.5, 0.3, 0.5),
  and (0.10, 0.43, 0.20, 0.27), respectively; the KS-test agreement
  with the full AMUSE distribution is $p=0.996$.} The expected nuclear
\hbox{X-ray} luminosities where an SMBH is present are calculated from
the $L_{\rm X}$--$M_{\rm star}$ correlation, here given by the
best-fit model to the full AMUSE sample (the solid black trend line),
but with significant intrinsic scatter to match that observed. Next,
each of the simulated galaxies is assigned an SMBH based on the choice
of $M_{\rm star,0}$; i.e., the high $M_{\rm star,0}$ red curve in
panel (b) populates only the high mass galaxies shown by the red
points in panel (a), whereas the intermediate $M_{\rm star,0}$ blue
curve populates the high mass galaxies down to the intermediate mass
galaxies shown by the red through blue points, and finally the low
$M_{\rm star,0}$ green curve populates nearly all galaxies down to
dwarfs shown by the red through green points. Panel (c) shows the
total simulated $M_{\rm star}$ distribution, which by construction
matches the AMUSE sample, and uses the same color coding to illustrate
the progression in occupation fraction as parameterized by $M_{\rm
  star,0}$. The conversion from simulated to observed $L_{\rm X}$ then
results from imposing a sensitivity threshold, such as the horizontal
dotted black line in panel (a) from AMUSE, and changing $L_{\rm X}$ to
an upper limit (with a value narrowly scattered around the threshold)
for all galaxies that either lack an SMBH or else have an SMBH
emitting below the detection sensitivity.

\begin{figure}
\includegraphics[scale=0.50]{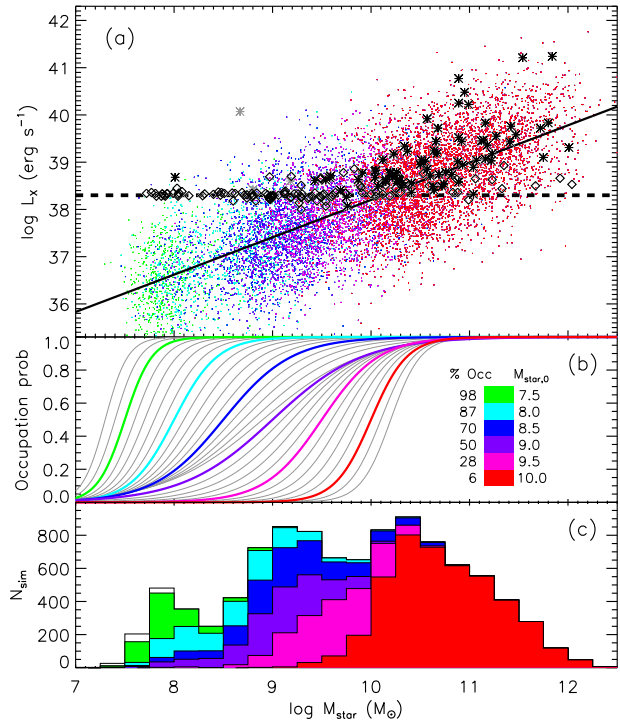} \figcaption{\small
  (a) X-ray luminosity versus stellar mass. AMUSE early-type galaxies
  are plotted as black stars (X-ray detections; gray stars are
  star-forming galaxies excluded from the clean sample defined in
  $\S$3) or as diamonds (upper limits). The horizontal dashed black
  line is the typical AMUSE sensitivity limit. The solid black trend
  line shows the best-fit relation for the full sample and the colored
  points are a random realization of this model. (b) Illustration of
  the parameterization used to model different occupation
  fractions. The colored lines show $\log{M_{\rm star,0}}$ values and
  their consequent occupation fraction (for galaxies with $\log{M_{\rm
      star}}<10$) as given in the legend; see text for details. (c)
  Distribution of $M_{\rm star}$ simulated from a sum of four
  Gaussians to match the AMUSE dataset; the histograms show occupied
  galaxies color-coded as in (b).}
\end{figure}

We modified the Bayesian linear regression code of Kelly (2007) to fit
for the SMBH occupation fraction (i.e., $M_{\rm star,0}$) while
simultaneously determining $L_{\rm X}$ as a function of $M_{\rm
  star}$. The primary difference between the model of Kelly (2007) and
our extension is that the method of Kelly (2007) would model the
distribution of $L_X$ at fixed $M_{\rm star}$ as a single normal
distribution, whereas we here model the distribution of $L_X|M_{\rm
  star}$ as a mixture of a normal distribuiton and a delta function
centered at an extremely small value of $L_X$, with the mixing weights
as a function of $M_{\rm star}$ given by the occupation fraction at
that $M_{\rm star}$. Specifically, we assume
\begin{eqnarray}
p(\log L_X|\log M_{\rm star}) = \nonumber \\ f_{\rm occ}(M_{\rm
  star})N(\log L_X| \alpha + \beta \log M_{\rm star}, \sigma^2)
\nonumber \\ + (1.0 - f_{\rm occ}(M_{\rm star})) \delta(\log L_X +
9999)
\end{eqnarray}
where $N(x|\mu,\sigma^2)$ denotes a normal distribution with mean
$\mu$ and variance $\sigma^2$ as a function of $x$, and $\alpha,
\beta,$ and $\sigma^2$ denote the intercept, slope, and variance in
the intrinsic scatter of the $\log L_X$--$\log M_{\rm star}$
relationship, and $\delta(\cdot)$ is the Dirac delta function.

In order to obtain samples of $\log M_{\rm star, 0}, \alpha, \beta,$
and $\sigma^2$ from their posterior distribution, we used an extension
of the Gibbs sampler of Kelly (2007). In our Gibbs sampler we
introduce a latent indicator variable, $I_i$, where $I_i = 1$ if the
$i^{\rm th}$ galaxy has a black hole in it and $I_i = 0$
otherwise. For all sources with X-ray detections $I_i = 1$ and is
considered known, while for those with upper limits $I_i$ is
unknown. For those sources with unknown $I_i$ we update their values
of $I_i$ at each stage of the Gibbs sampler by drawing from a
Bernoulli distribution with probability
\begin{eqnarray}
p(I_i=1|L_{X,i}, M_{\rm star, i}, \alpha, \beta, \sigma^2, M_{\rm
  star, 0}) = \nonumber \\ \Phi(\frac{\log L_{X,i} - \alpha - \beta
  \log M_{\rm star, i}}{\sigma}) \nonumber \\ \left[ 1 - (1 +
  \Phi(\frac{\log L_{X,i} - \alpha - \beta \log M_{\rm star,
      i}}{\sigma})) f_{\rm occ}(M_{\rm star, 0}) \right]^{-1}
\end{eqnarray}
where $\Phi(\cdot)$ denotes the cumulative distribution function of
the standard normal distribution. Given these new values of $I_i$ we
can then update $M_{\rm star, 0}$ using a Metropolis update in
combination with the conditional posterior
\begin{eqnarray}
p(\log M_{\rm star, 0}|I_1,\ldots,I_n) = \nonumber \\ \prod_{i=1}^n
[f_{\rm occ}(M_{\rm star,0})]^{I_i} [1 - f_{\rm occ}(M_{\rm
    star,0})]^{1-I_i}.
\end{eqnarray}
The rest of the Gibbs sampler proceeds as in Kelly (2007).

The parameter $\log{M_{\rm star,0}}$ is restricted to be greater than
7.5 since any values below 7.5 already produce near 100\% occupation
fraction. As with the original {\tt linmix\_err} IDL routine,
measurement errors, intrinsic scatter, and upper limits are
incorporated, and the independent variable distribution is
approximated as a sum of Gaussians. The four parameters of interest in
our model are the intercept, slope, and intrinsic scatter of the
$L_{\rm X}(M_{\rm star})$ relation as well as $\log{M_{\rm star,0}}$,
which gives the occupation fraction for galaxies below $M_{\rm
  star}=10^{10}~M_{\odot}$. The best-fit preferred parameter values
are taken as the median of 5000 (thinned from 50000, retaining every
tenth) draws from the posterior distribution and quoted errors
correspond to 1$\sigma$ uncertainties.

\begin{figure}
\includegraphics[scale=0.52]{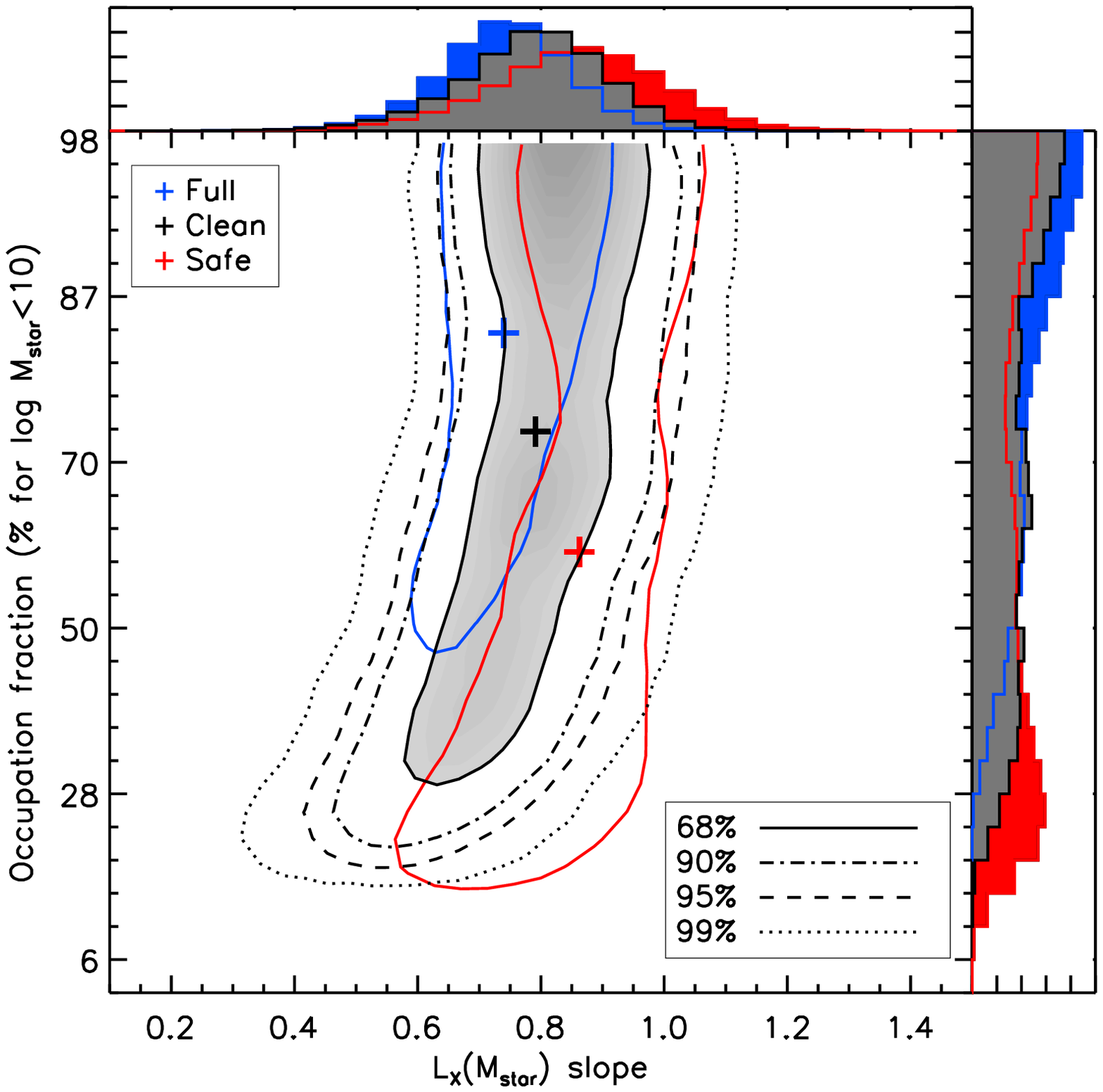}
\includegraphics[scale=0.50]{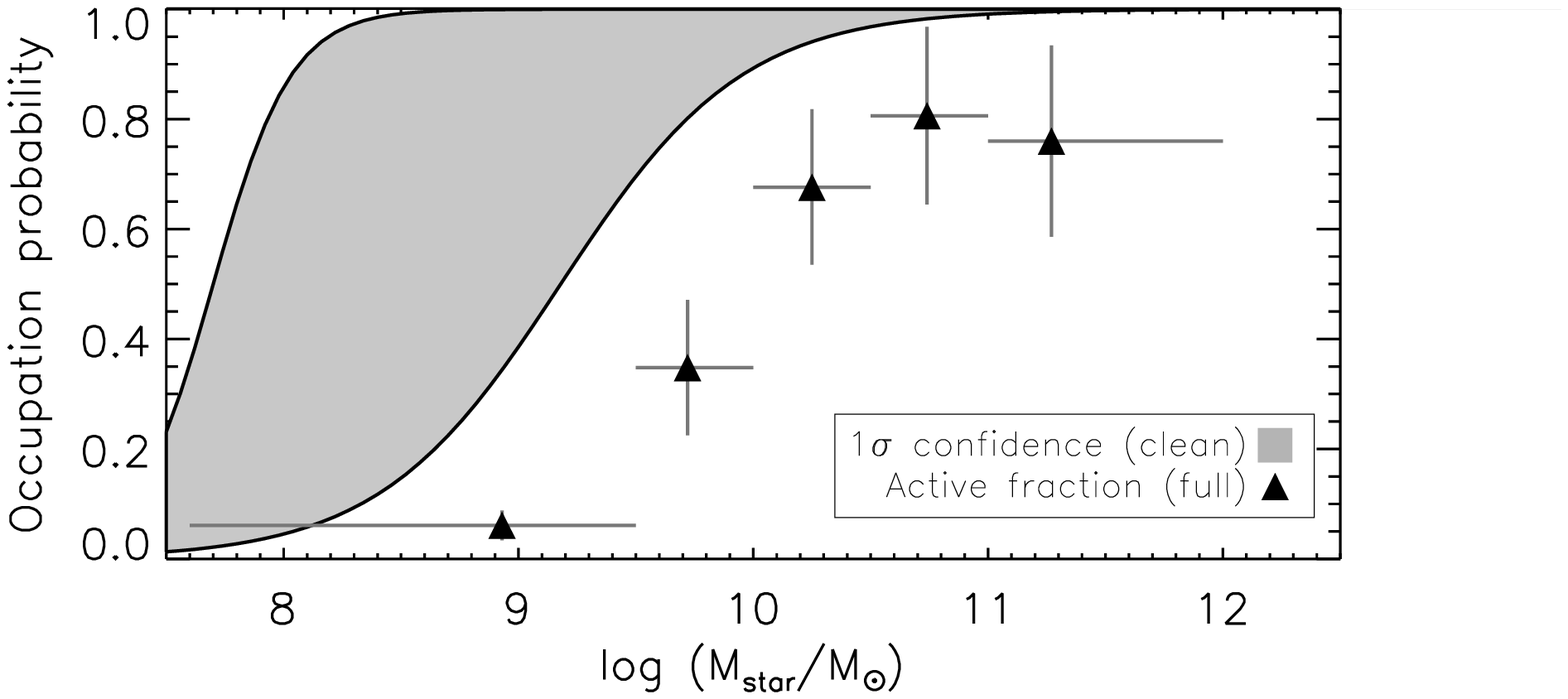} \figcaption{\small {\it
    Top:\/} Preferred model for the AMUSE dataset, for the clean,
  full, and safe samples (black, blue, and red; see text for
  details). The median values from the posterior probability are
  marked with crosses, and the histograms show the marginalized
  distributions. Joint 68\%, 90\%, 95\%, and 99\% confidence contours
  are plotted for the clean sample. {\it Bottom:\/} Permitted
  occupation probability (1$\sigma$ confidence) as a function of
  $M_{\rm star}$ for the clean AMUSE dataset, along with the active
  fraction from the full sample. The active fraction provides a lower
  limit for the occupation fraction, and the full sample provides the
  highest detection fraction.}
\end{figure}

\section{Results from the AMUSE surveys}

Stellar masses and X-ray luminosities for the AMUSE Virgo and Field
sample were previously published in G10 and M12a.\footnote{NGC 5077
  ($d>30$~Mpc) and NGC 4627 (atypically deep serendipitous coverage)
  are here removed from the Field sample.} As described in those
works, the detected nuclear {\it Chandra\/} \hbox{X-ray} sources are
point-like and located at the projected optical center of their
galaxy, to within the optical and \hbox{X-ray} astrometric and
centroid uncertainties. We determine more precise stellar masses for
some of the AMUSE galaxies using archival and newly obtained {\it
  HST\/} data, including Cycle 19 two-color {\it HST\/} ACS imaging of
Field galaxies (Baldassare et al.~2014; B14 hereafter). Two Field
galaxies, NGC 3928 (described as a starburst by Carollo et al.~1997)
and NGC 3265, show spiral arms in {\it HST\/} imaging and are
removed. We also cautiously choose to set aside VCC 1857 and VCC 1828
from the Virgo sample, as these two galaxies have irregular and late
types, respectively, in HyperLeda despite their arguably elliptical
appearance. The ``full'' AMUSE sample then consists of 197 early-type
galaxies of which 81 (or 41\%) have \hbox{X-ray} detections.

The distances to the Virgo galaxies were assumed as 16.4 Mpc in G10
and the distances to the Field galaxies were calculated from their
redshifts in M12a. We here (and in B14) make use of slightly more
accurate distances, specifically taking distances to Virgo galaxies
from Mei et al.~(2007) and distances to Field galaxies from
non-redshift measurements given in HyperLeda where available. The
stellar masses and X-ray luminosities are adjusted from G10 and M12a
using these more accurate distances (with {\it HST\/}-derived $M_{\rm
  star}$ values for some Field galaxies taken from B14); the resulting
full AMUSE Virgo and Field combined sample properties are given in
Table~1. While the properties for several individual galaxies are
improved in accuracy, this adjustment has only a tiny statistical
impact on the overall sample, with a median change to $M_{\rm star}$
and $L_{\rm X}$ of 0.02 dex (standard deviation of 0.11 dex).

We next generate a ``clean'' sample by removing three galaxies for
which optical or UV {\it HST\/} imaging shows irregular or clumpy
morphology and colors suggestive of recent ($<$100-300~Myr) star
formation. From the Field sample, NGC~855 and ESO~540$-$014 display
clumpy structure with blue colors (a spectrum of the latter indicates
high star-formation rates rather than the literature Seyfert 2
classification; Amy Reines, private communication). From the Virgo
sample, VCC 1499 has archival {\it HST\/} UV imaging that is
suggestive of galaxy-wide star-formation, and it is identified as
post-starburst by Gavazzi et al.~(2001). These three galaxies all have
detectable nuclear \hbox{X-ray} emission, but their optical
morphologies and blue colors indicate \hbox{X-ray} contamination from
high-mass \hbox{X-ray} binaries is possible. The clean sample then
contains 194 galaxies of which 78 have \hbox{X-ray}
detections.\footnote{We reconfirm using the clean sample the
  marginally significant finding from M12b that the Field galaxies
  tend to be X-ray brighter than their Virgo counterparts.}

Finally, we generate a ``safe'' sample by correcting for potential
contamination of the nuclear \hbox{X-ray} emission due to low-mass
\hbox{X-ray} binaries (LMXBs). The probability of contamination of the
nuclear X-ray emission from LMXBs enclosed by the projected {\it
  Chandra\/} point spread function, within the $d<30$~Mpc
volume-limited AMUSE sample, is generally negligible for non-nucleated
galaxies (see G10 for details; for context, $1''$ corresponds to
projected 50/100/150~pc at distances of 10/20/30~Mpc). For galaxies
hosting a nuclear star cluster, the probability of contamination is
greater and is conservatively estimated using a globular cluster
\hbox{X-ray} luminosity function (G10; B14). Taking into account the
measured nuclear \hbox{X-ray} luminosities, the probability of LMXB
contamination is $\simgt$50\% in four Virgo galaxies (VCC~1883,
VCC~784, VCC~1250, and VCC 1283 have contamination probabilities of
100\%, 51\%, 100\%, and 45\%; G10). From the {\it HST\/}-covered Field
sample NGC~3384, NGC~1172, and NGC~2970 have non-negligible
contamination probabilities (B14); NGC~3384 is known from stellar
dynamics to have an SMBH with a mass of $\sim$1.8$\times10^{7}
M_{\odot}$ (Gebhardt et al.~2000) so we flag NGC~1172 and
NGC~2970. NGC 1331 is the only additional Field galaxy known to have
both a NSC and a central X-ray source, but it has a $<$5\% probability
of LMXB contamination. From among the Field galaxies that have a
detected nuclear \hbox{X-ray} source but lack {\it HST\/} coverage,
about three more are statistically expected to contain nuclear star
clusters that could generate LMXB contamination. An example is
NGC~3522 with $\log{M_{\rm star}/M_{\odot}}<10$ and a moderate
$\log{L_{\rm X}}=38.8$, and we randomly chose two others from
NGC~2778, NGC~3457, NGC~3641, NGC~4283, NGC~1370, or PGC~1206166 in
four ways, producing four slightly different versions of a safe
sample. The \hbox{X-ray} luminosities for the nine flagged galaxies
are converted to limits for the safe sample, which then still contains
194 galaxies but with 69 now considered to be \hbox{X-ray}
detections. The four slightly different versions of the safe sample
produce formally consistent fitting results, although the median
occupation fraction is lower if PGC~1206166 with $\log{M_{\rm
    star}/M_{\odot}}=8.0$ is considered contaminated (illustrating the
importance of every X-ray detection in the dwarf galaxy regime).

The safe sample provides a deliberately cautious approach to LMXB
contamination. Both the Virgo and Field samples include several
galaxies with nuclear star clusters that are not detected in
\hbox{X-rays}; for example, from the Field NGC~1340 and NGC~1426 have
calculated probabilities of $\sim$10\% for having a central LMXB with
$L_{\rm X}$ greater than the AMUSE sensitivity. Because the
probability of hosting a nuclear star cluster increases to lower
stellar mass (as does the profile cuspiness, centrally partially
offsetting the absolute decrease in $M_{\rm star}$), the potentially
contaminated galaxies all have $M_{\rm star}<3\times10^{10}
M_{\odot}$. (The Field sample has relatively more low-mass galaxies,
nucleated galaxies, and potentially contaminated galaxies, but the
frequency of nuclear star clusters after accounting for stellar mass
is similar to that in Virgo; B14). The impact of conservatively
correcting for LMXB contamination is to reduce slightly the inferred
occupation fraction and to increase slightly the slope. The most
accurate representation of the AMUSE dataset likely lies between the
clean and safe samples, probably closer to the former given the
conservative LMXB correction.
 
A low percentage of \hbox{X-ray} detections in low-mass galaxies
arises from some combination of a low occupation fraction, a steep
$L_{\rm X}(M_{\rm star})$ slope, and small intrinsic scatter. For the
AMUSE sample, the overall detection fraction is 41\% (36\% after
accounting for potential LMXB contamination); most of these detections
(84\%, or slightly higher after LXMB correction) are in galaxies with
$\log{M_{\rm star}}>10$, and only 1--2\% of galaxies with $\log{M_{\rm
    star}}<9$ have nuclear \hbox{X-ray} detections. For illustration,
similar distributions can be produced for slopes of 1.0 with
occupation fractions between 50\% and 100\% and intrinsic scatter of
0.7 dex, or for slopes of 0.7 with 50\% occupation fraction and
intrinsic scatter of 0.5 dex. However, flatter slopes $<0.4$ that
match the overall detection fraction necessarily overpredict the
proportion of detections in low-mass galaxies, and occupation
fractions of $<$20\% underpredict that same ratio.

The results of applying this modeling framework to the AMUSE dataset
are shown in Figure~2. The posterior distributions of the slope and
the occupation fraction for the full, clean, and safe AMUSE samples
are plotted as confidence contours and marginalized histograms. (To
illustrate the spread in the safe samples we plot all four versions
combined.) The slope is relatively well-constrained even with
occupation fraction as a free parameter (0.74$\pm$0.10, 0.79$\pm0.12$,
or 0.86$\pm$0.14 for the full, clean, and safe samples). For the clean
sample, the slope has a negligible ($\simlt$0.05) probability of being
$<$0.5 or $>$1.0. However, the occupation fraction is only loosely
constrained; the probability distribution extends from 30\% to 100\%
($p=0.34, 0.46$ for occupation $>87$\%, $<70$\%). Only occupation
fractions of $<$20\% are securely ruled out by our data.  (Recall that
dynamical SMBH mass limits for M33 and NGC~205 argue against a 100\%
occupation fraction). Figure~2, bottom, shows the 1$\sigma$ confidence
region for the occupation probability as a function of $M_{\rm star}$,
along with the lower limits provided by the X-ray active fraction. The
significant uncertainties prevent a definitive discrimination between
formation mechanisms.\footnote{Since both pathways are theoretically
  viable and presumably operate at some level, definitive
  identification of the dominant seeding mode would not rule out that
  some supermassive black holes formed from alternative mechanisms.}

The preferred $L_{\rm X}(M_{\rm star})$ slope of $\sim$0.7--0.8 for
the full or clean AMUSE samples supports downsizing in these weakly
accreting SMBHs, albeit with respect to $M_{\rm star}$ rather than
inferred $M_{\rm BH}$ as given in G10 and M12b. (We used simulations
to confirm that an input universal Eddington ratio distribution would
produce preferred slopes $\simeq1$ with our methodology.)  While
dynamical measurements of $M_{\rm BH}$ in these galaxies are not
available, a typical scaling of $M_{\rm BH}{\sim}M_{\rm
  bulge}^{1.12\pm0.06}$ (H{\"a}ring \& Rix~2004) would suggests this
downsizing effect is similar or perhaps slightly more pronounced with
black hole mass rather than host galaxy stellar mass.\footnote{For
  most of these early-type galaxies $M_{\rm bulge} \simeq M_{\rm
    star}$; also, applying a bulge-to-total correction would make the
  downsizing more extreme.} Additional dynamical measurements (such as
that in NGC~404; Seth et al.~2010; see also Neumayer \& Walcher 2012)
are required to confirm the intriguing apparent tendency for $M_{\rm
  BH}$ in lower mass galaxies to fall below the extrapolation of the
H{\"a}ring \& Rix~(2004) relation (see Greene 2012 and references
therein). It seems unlikely that $M_{\rm BH}/M_{\rm star}$ could
increase for dwarfs such that $\log{L_{\rm X}}$ could scale linearly
with $\log{M_{\rm BH}}$.

We emphasize that these results are derived from early-type galaxies,
and insofar as SMBH seeding and growth is linked to bulge properties
rather than stellar mass (e.g., Beifiori et al.~2012) they may not
apply to late-type or irregular galaxies.

\begin{figure}
\includegraphics[scale=0.57]{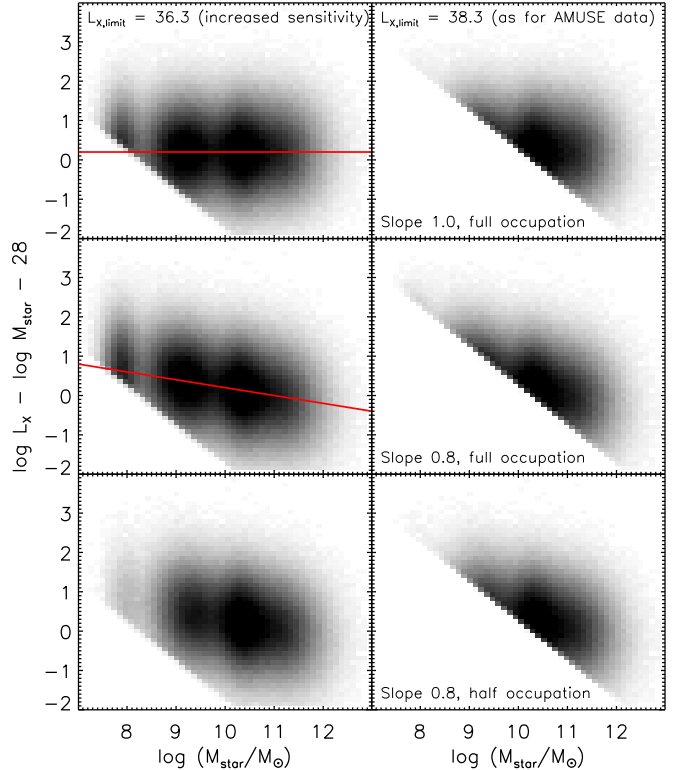} \figcaption{\small
  Simulated distribution of X-ray detected objects for an artificially
  increased sensitivity of $L_{\rm X}=36.3$ (left column) and for the
  AMUSE sensitivity of $L_{\rm X}=38.3$ (right column). 10$^{6}$
  points are binned in a 50-by-50 tiling, and the density is plotted
  in grayscale with squareroot scaling. The top and middle rows are
  for full occupation with slopes of 1.0 (uniform Eddington
  efficiency) and 0.8 (downsizing), while the bottom row is for an
  occupation fraction of $\sim$50\% for $M_{\rm star}<10^{10}
  M_{\odot}$, again with a downsizing 0.8 slope.}
\end{figure}

\section{Assessing uncertainties and future prospects from simulations}

To assess prospects for improving constraints upon the occupation
fraction with future surveys, we use the Bayesian linear regression
fitting to investigate the impact of the limiting sensitivity and the
sample size upon the parameter errors.

\begin{figure*}
\includegraphics[scale=0.51]{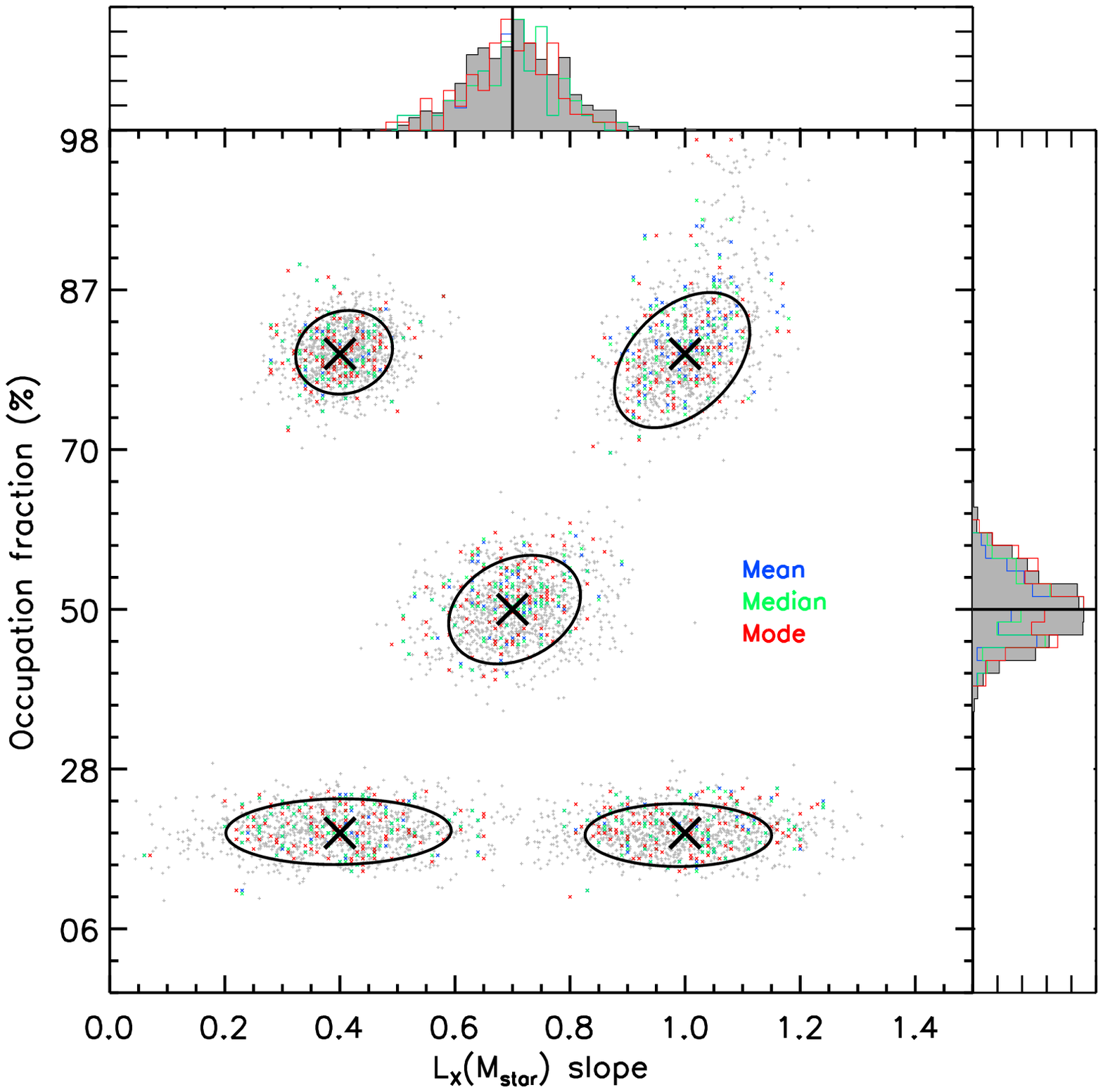}
\hspace{0.4cm} \includegraphics[scale=0.51]{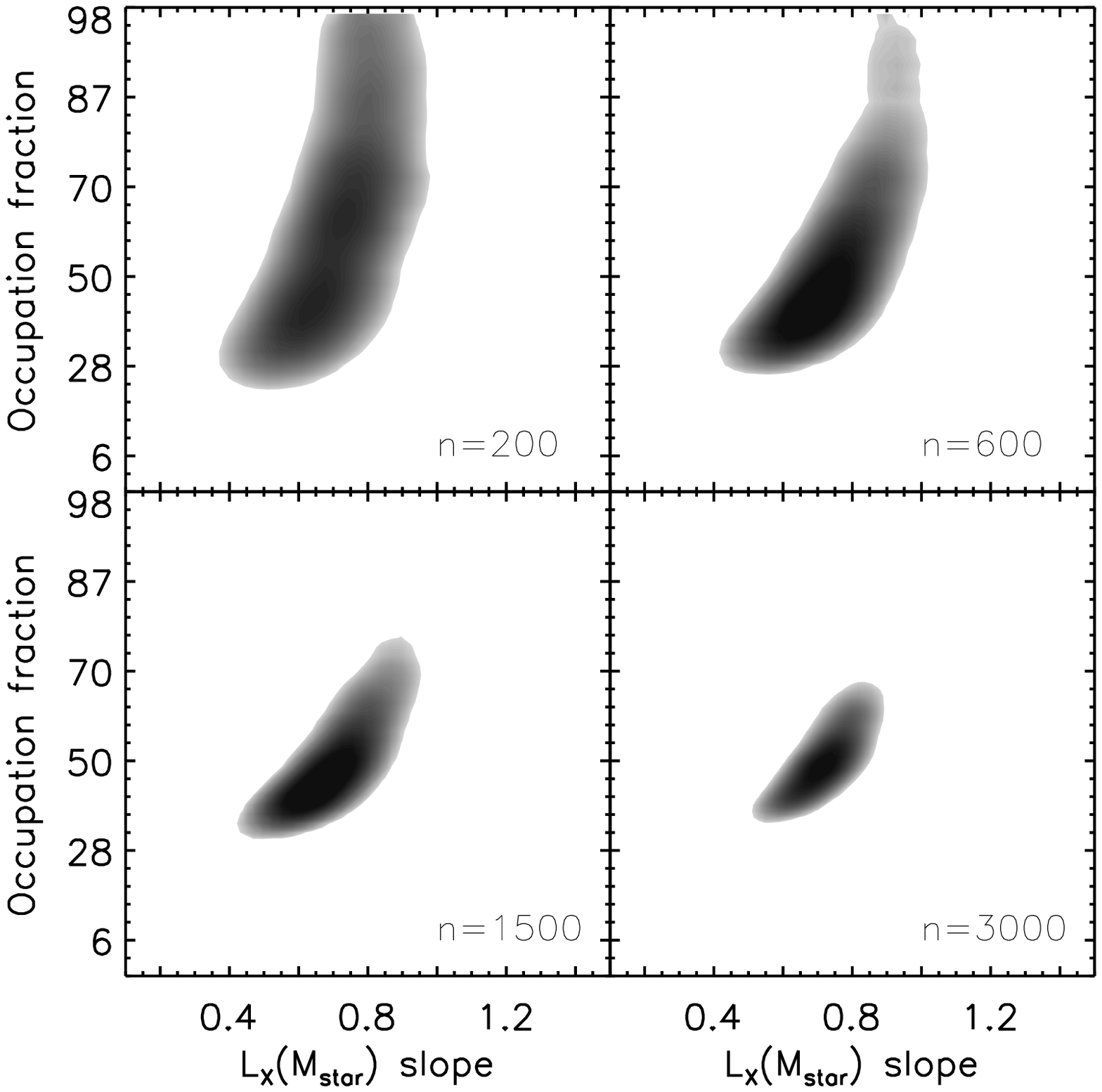}
\figcaption{\small {\it Left\/}: Illustration of the uncertainties in
  the slope and occupation fraction for simulations from different
  input models for an artificially increased sensitivity $\log{L_{\rm
      X,limit}}=36.3$~erg~s$^{-1}$. The input parameters are cleanly
  recovered with only 200 simulated data points.  {\it Right\/}:
  Illustration of the uncertainties in the slope and occupation
  fraction for simulations with differing sample sizes, as indicated,
  for sensitivity $\log{L_{\rm X,limit}}=38.3$~erg~s$^{-1}$ matching
  the AMUSE dataset.}
\end{figure*}

The distribution of $L_{\rm X}$ versus $M_{\rm star}$ is only subtly
changed by partial occupation fraction for the AMUSE sensitivity limit
(Figure~3, right column), because most of the impacted low-mass
galaxies would already have X-ray luminosities precluding
detection. To examine the impact of the sensitivity limit (as well as
to validate our modeling techniques), we also consider an artificially
increased sensitivity of $\log{L_{\rm X,limit}}=36.3$~erg~s$^{-1}$,
two orders of magnitude below that for the AMUSE surveys. This
contrived model usefully illustrates the impact of downsizing and
partial occupation (Figure~3, left).

We verified that arbitrary input parameters are cleanly recovered
(with correct statistical uncertainties) in simulations given an
artificially increased sensitivity of $\log{L_{\rm
    X,limit}}=36.3$~erg~s$^{-1}$. Recall that for these simulations
$M_{\rm star}$ values are drawn from a sum of four Gaussians which
empirically matches the AMUSE distribution, $L_{\rm X}$ is computed
from $M_{\rm star}$ for a given correlation, and then the occupation
fraction is enforced following the probability curve for a given
$M_{\rm star,0}$ value (Figure~1b) and objects then lacking an SMBH are
mandated to be \hbox{X-ray} upper limits. Some examples of fitting
these simulations are provided in Figure~4 for input slopes of 0.4,
0.7, and 1.0 and occupation fractions of 15\%, 50\%, and 85\%. The
sample size was fixed to 200 objects, and the simulated points were
varied by an intrinsic $L_{\rm X}(M_{\rm star})$ scatter of 0.7 dex
for 100 realizations of each model. The resulting uncertainties on
both parameters are modest even with only 200 points (and consistent
with the output errors from the code). Unfortunately, even the
outstanding {\it Chandra\/} PSF is not sufficient to overcome the
rapid rise in the luminosity function of low-mass \hbox{X-ray}
binaries and so contamination becomes impossible to avoid below the
AMUSE sensitivity limit of $\log{L_{\rm
    X}}\sim38.3$~erg~s$^{-1}$. With higher spatial resolution the
total projected stellar mass enclosed in an \hbox{X-ray} extraction
aperture, and correspondingly the likelihood of contamination, could
be decreased. To achieve a 20$\times$ improvement in sensitivity down
to $\log{L_{\rm X, limit}}\sim37.0$~erg~s$^{-1}$, for galaxies of
stellar mass $\sim10^{9} M_{\odot}$ lying within 30 Mpc (with
effective radii of $\sim15''$) a resolution of $\sim0.05''$ would be
necessary to limit potential contamination to $<10$\% in a given
\hbox{X-ray} nuclear detection. The inclusion of X-ray variability or
spectral measurements (or other activity indicators, such as radio
emission) could weaken this requirement.

The impact of increasing the sample size is also shown in Figure~4,
for $\log{L_{\rm X,limit}}=38.3$~erg~s$^{-1}$ as for the AMUSE
dataset. For these simulations new objects are added at the AMUSE
$M_{\rm star}$ distribution probabilities but weighted by a factor of
two for $\log{M_{\rm star}}<10$; the uncertainties on the occupation
fraction converge more quickly when smaller galaxies are
preferentially targeted. With only 600 total objects, the statistical
errors on the occupation fraction permit clean differentiation between
full and half occupation, and with 1500 objects the occupation can be
fixed to $\pm$15\%. If the slope or intrinsic scatter were known (for
example, from theoretical arguments) rather than fit, far fewer
objects would be required to obtain such constraints. There are good
prospects for combining the AMUSE samples with new coverage (e.g., of
the outer Fornax cluster), or with archival coverage of low-mass
galaxies serendipitously present in existing very deep {\it Chandra\/}
observations of M87 in Virgo or 3C~84 in Perseus. Here ultra-compact
dwarf galaxies, which may contain SMBHs and have undergone tidal
stripping (Mieske et al.~2013; Seth et al.~2014), could be included.

\section{Discussion and future applications}

Our simultaneous fitting of the SMBH occupation fraction and the
scaling of nuclear \hbox{X-ray} luminosity with stellar mass
constrains SMBHs to be present in $>$20\% of early-type galaxies with
$M_{\rm star}<10^{10} M_{\odot}$ and suggests the dependence of
$\log{L_{\rm X}}$ upon $\log{M_{\rm star}}$ has a slope of
$\sim$0.7--0.8. This work provides promising if inconclusive
information on the local SMBH occupation fraction and also supports a
downsizing trend in low-level SMBH activity. 

The highly sub-Eddington objects that make up the AMUSE dataset are
expected to feature radiatively inefficient accretion flows
(RIAFs). Bondi accretion of even the limited gas provided by stellar
winds (Volonteri et al.~2011) near the nuclei of early-type galaxies
would predict greater \hbox{X-ray} luminosities than observed; the
efficiency as well as the accretion rate must be low in these objects
(Soria et al.~2006; Ho 2009). Either an advection-dominated accretion
flow (e.g., Di Matteo \& Fabian 1997; Narayan et al.~1998) or an
outflow/jet component (e.g., Soria et al.~2006; Plotkin et al.~2012)
is required. In general, the efficiency in these hot flows is
theoretically expected to decrease toward lower accretion rates (Yuan
\& Narayan 2014 and references therein). Although the Bondi radius is
directly resolved by Chandra in deep observations of NGC~3115, the
temperature profile is inconsistent with simple RIAF models (Wong et
al.~2014). Fueling of a RIAF by steady-state stellar winds may be
supplemented by intermittent processes such as tidal disruption, or by
gradual stripping of central stars (e.g., MacLeod et al.~2013). While
we cannot constrain the physical mechanism responsible for the
observed \hbox{X-ray} emission, the simplest explanation for
downsizing in low-level SMBH activity would be that the relative rate
of accretion is higher in smaller galaxies, with a fixed low
efficiency. We reiterate that the downsizing we identify here is
restricted to low-level SMBH activity and may not apply to AGNs.

The methodology we use here is flexible and could also be applied to
deep surveys of AGN. For example, the 4~Ms CDFS contains active
galactic nuclei including to relatively modest $M_{\rm
  star}\simlt3\times10^{9} M_{\odot}$ (Schramm \& Silverman 2013;
Schramm et al.~2013) and at low levels of activity (Young et al.~2012)
as well as normal galaxies at cosmological distances (Lehmer et
al.~2012), and opens substantial additional volume albeit at lower
sensitivity. We provide an illustration of applying this general
technique to simulated deep field galaxies in Figure~5, and are
pursuing this approach in Greene et al.~(in preparation).

In this higher $L_{\rm X}$ regime we populate X-ray luminosities
drawing from a uniform Eddington ratio distribution with a power-law
slope of $-0.65$ as in Aird et al.~(2012), assuming $\log{M_{\rm BH}}
= \log{M_{\rm star}}-2.8$. For a hypothetical combined
CDFS+CDFN+AEGISXD sample with a typical detection sensitivity of
$\log{L_{\rm X}}\simeq40$ out to $z<0.4$, we expect about 15000
$z<0.4$ galaxies (estimated from Cardamone et al.~2010; Xue et
al.~2010) of which $\sim$300 or 2\% should host \hbox{X-ray} AGNs
(estimated from Xue et al.~2010, 2011; Lehmer et al.~2012). We
confirmed with an artificially large sample that the distribution of
\hbox{X-ray} detections (color gradients in Figure~5) can be used to
infer the occupation fraction. For example, the percentages of
\hbox{X-ray} AGNs in hosts with $\log{M_{\rm star}}<9.5$ is 11.9\%
with full occupation versus 6.1\% with half occupation. This is
statistically distinguishable at 99\% confidence for the expected
$\simeq$300 AGNs (black crosses in Figure~5) if the other model
parameters are known or fixed by theory; full versus half occupation
predicts 37 versus 17 \hbox{X-ray} AGNs in hosts with $\log{M_{\rm
    star}}<9.5$). This test will increase in power with the upcoming
deeper CDFS exposure.

We are also refining our measurement of occupation fraction within the
AMUSE sample through incorporating the influence of large-scale
environment upon low-level SMBH activity. The scaled nuclear
\hbox{X-ray} luminosities of early-type galaxies apparently decrease
from isolated to group to cluster environments (M12a,b). This may
reflect greater quantities of cold gas in field galaxies (e.g.,
Oosterloo et al.~2010), for example due to reduced stripping relative
to their cluster counterparts. Cold accretion has been inferred to be
relevant to low-level SMBH activity from studies of brightest cluster
galaxies (Russell et al.~2013) and dust (Martini et al.~2013), and
AGNs preferentially inhabit gas-rich galaxies (Vito et al.~2014). The
recent tentative finding that nuclear star clusters in massive early
type galaxies are bluer in the field (B14) implies that field nuclear
star clusters formed at lower metallicities and/or experienced more
recent star formation, relative to cluster counterparts; this in turn
suggests that cold gas can eventually filter down to the central
regions where it is available (either directly or via enhanced star
formation and stellar winds) to be heated and inefficiently accreted
onto the central SMBH. We are continuing to investigate the impact of
Mpc-scale densities using new {\it Chandra\/} observations of
early-type galaxies located within cosmic voids. However, the analysis
presented here is not biased because the slopes of the $L_{\rm
  X}-M_{\rm star}$ relation are consistent between the AMUSE-Field and
AMUSE-Virgo samples (M12b); instead, the uncertainties are potentially
slightly inflated. Including any environmental dependence, once
quantified at high significance, will helpfully decrease the scatter
in the $L_{\rm X}(M_{\rm star})$ relation in the combined AMUSE
dataset.

Additional multiwavelength information will provide better
understanding of both individual objects and of the overall population
(e.g., the distribution of galaxies showing radio, or optical,
indications of nuclear activity; Reines et al.~2013). New dynamical
mass measurements with a 30m class telescope would help clarify the
mass distribution of SMBHs in smaller galaxies, providing a
complementary probe of black hole birth and growth (van Wassenhove et
al.~2010).  In this context it is interesting that no galaxies with
$M_{\rm star}<10^{10} M_{\odot}$ (without stripping; Seth et al.~2014)
are yet known with confirmed $M_{\rm BH}>10^{6} M_{\odot}$. Tidal
disruption transients, particularly from white dwarfs, can provide
complementary insight into lower-mass SMBHs (Clausen \& Eracleous
2011; MacLeod et al.~2014). Pairing observational advances with
increasingly sophisticated theoretical models will help discriminate
between models of seed formation.
 
\begin{figure}
\includegraphics[scale=0.68]{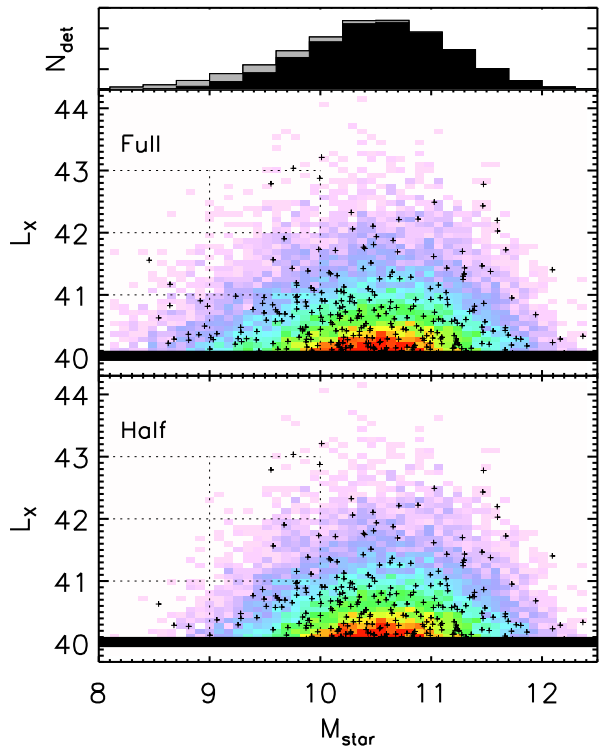} \figcaption{\small
  Distribution of AGN X-ray detections for mock deep field catalogs
  with 50\% and 100\% occupation fractions for $M_{\rm star}<10^{10}
  M_{\odot}$. The colors indicate detection density with the black
  crosses a realization with 15000 total galaxies and $\sim$300 X-ray
  AGN, $\pm$20 depending on the occupation fraction. The top histogram
  shows X-ray detected AGN for half (black) and full (gray)
  occupation.}
\end{figure}

\acknowledgments

We thank Andy Fabian, Rich Plotkin, Amy Reines, Claudia Scarlata, and
Anil Seth for helpful discussions, and an anonymous referee for
constructive comments. This work was supported in part by Chandra
Award Number 11620915, by the National Science Foundation under grant
no.~NSF PHY11-25915 and by NASA through grant HST-GO-12591.01 from the
Space Telescope Science Institute, which is operated by AURA, Inc.,
under NASA contract NAS 5-26555. JHW acknowledges support by the
National Research Foundation of Korea (NRF) grant funded by the Korea
government (MEST; No.~2012-006087).

\clearpage
\LongTables
\begin{deluxetable*}{lrrrrrr}
\tabletypesize{\footnotesize}
\tablecolumns{7}
\tablewidth{0pt}
\tablecaption{Combined AMUSE Virgo and Field sample of early-type galaxies}

\tablehead{ \colhead{Name} & \colhead{V/F} & \colhead{Distance} &
  \colhead{Method} & \colhead{$\log{M_{\rm star}}$} &
  \colhead{$\log{L_{\rm X}}$} & \colhead{Notes} \\ & & (Mpc) & &
  ($M_{\odot}$) & (erg~s$^{-1}$) & \\ & & & & & & }

\startdata
                  VCC 1226 & V & 17.1 & M07 & 12.0 & $<$38.5 &               \\ 
                   VCC 731 & V & 23.3 & M07 & 12.0 &  39.3   &               \\
                   VCC 881 & V & 16.8 & M07 & 11.9 & $<$38.7 &               \\
                  VCC 1316 & V & 17.2 & M07 & 11.8 &  41.2   &               \\
                   VCC 763 & V & 18.4 & M07 & 11.8 &  39.8   &               \\
                  VCC 1978 & V & 17.3 & M07 & 11.8 &  39.1   &               \\
                  NGC 1407 & F & 28.6 &  HL & 11.7 &  39.7   &               \\
                   VCC 798 & V & 17.9 & M07 & 11.7 & $<$38.5 &               \\
                   IC 1459 & F & 29.0 &  HL & 11.5 &  41.2   &               \\
                  NGC 5322 & F & 30.9 &  HL & 11.5 &  39.6   &               \\
                  NGC 2768 & F & 22.2 &  HL & 11.4 &  39.8   &               \\
                  NGC 0720 & F & 27.4 &  HL & 11.4 &  39.4   &               \\
                  NGC 5846 & F & 24.7 &  HL & 11.3 & $<$38.9 &               \\
                  NGC 3923 & F & 20.0 &  HL & 11.3 & $<$38.4 &               \\
                  VCC 1632 & V & 15.8 & M07 & 11.3 &  39.5   &               \\
                  NGC 7507 & F & 24.8 &  HL & 11.2 &  39.2   &               \\
                  NGC 3640 & F & 26.8 &  HL & 11.2 & $<$38.6 &               \\
                  VCC 1903 & V & 14.9 & M07 & 11.2 &  39.0   &               \\
                  NGC 1332 & F & 22.7 &  HL & 11.2 &  39.4   &               \\
                  NGC 4125 & F & 23.7 &  HL & 11.2 &  39.1   &               \\
                  NGC 4494 & F & 16.7 &  HL & 11.2 &  39.8   &               \\
                  NGC 3610 & F & 32.5 &  HL & 11.1 &  39.5   &               \\
                  NGC 3193 & F & 33.7 &  HL & 11.1 &  39.5   &               \\
                  NGC 3585 & F & 19.9 &  HL & 11.1 &  39.0   &               \\
                   VCC 575 & V & 22.1 & M07 & 11.1 & $<$38.6 &               \\
                  NGC 0821 & F & 23.3 &  HL & 11.0 &  38.9   &               \\
                  NGC 4636 & F & 14.1 &  HL & 11.0 &  38.3   &               \\
                  VCC 1535 & V & 16.3 &  HL & 11.0 & $<$38.2 &               \\
                  NGC 4036 & F & 21.1 &  HL & 11.0 &  40.2   &               \\
                  NGC 1052 & F & 17.5 &  HL & 10.9 &  40.5   &               \\
                  NGC 5576 & F & 25.2 &  HL & 10.9 &  38.9   &               \\
                  NGC 5838 & F & 19.5 &   z & 10.9 &  39.4   &               \\
                  VCC 2092 & V & 16.1 & M07 & 10.9 &  38.6   &               \\
                  NGC 4291 & F & 32.2 &  HL & 10.9 &  39.5   &               \\
                  NGC 4278 & F & 18.5 &  HL & 10.9 &  40.2   &               \\
                  NGC 4203 & F & 15.0 &  HL & 10.9 &  40.8   &               \\
                  NGC 5638 & F & 26.1 &  HL & 10.9 & $<$38.4 &               \\
                  VCC 1154 & V & 16.1 & M07 & 10.9 &  39.0   &               \\
                  NGC 1340 & F & 20.6 &  HL & 10.9 & $<$38.6 &               \\
                   VCC 759 & V & 17.0 & M07 & 10.8 & $<$38.4 &               \\
                  VCC 1030 & V & 16.8 & M07 & 10.8 &  38.7   &               \\
                  NGC 4697 & F & 12.2 &  HL & 10.8 &  38.8   &               \\
                  VCC 1231 & V & 15.3 & M07 & 10.7 &  38.5   &               \\
                  NGC 3379 & F & 11.3 &  HL & 10.7 &  38.5   &               \\
                  NGC 3115 & F &  9.6 &  HL & 10.7 &  38.7   &               \\
                  NGC 5845 & F & 32.7 &  HL & 10.7 &  39.7   &               \\
                  VCC 1025 & V & 22.4 & M07 & 10.7 &  39.2   &               \\
                  NGC 5831 & F & 26.9 &  HL & 10.7 &  39.4   &               \\
                  NGC 1439 & F & 26.4 &  HL & 10.6 &  39.2   &               \\
                  VCC 1062 & V & 15.3 & M07 & 10.6 &  38.4   &               \\
                  VCC 1692 & V & 17.1 & M07 & 10.6 &  38.5   &               \\
                  VCC 2095 & V & 16.4 & Vir & 10.6 &  38.7   &               \\
                  NGC 1426 & F & 23.3 &  HL & 10.6 & $<$38.5 &               \\
                  VCC 1664 & V & 15.8 & M07 & 10.6 &  39.9   &               \\
                  NGC 5582 & F & 28.2 &  HL & 10.6 &  38.9   &               \\
                  VCC 1938 & V & 17.5 & M07 & 10.6 &  39.0   &               \\
                  VCC 1279 & V & 17.0 & M07 & 10.5 & $<$38.8 &               \\
                   VCC 685 & V & 14.9 &  HL & 10.5 &  39.0   &               \\
                  NGC 4648 & F & 25.4 &   z & 10.5 &  39.0   &               \\
                  VCC 1883 & V & 16.6 & M07 & 10.4 &  38.4   &  NSC + X-ray  \\
                  NGC 3384 & F &  9.4 &  HL & 10.4 &  38.6   &  NSC + X-ray  \\
                  VCC 1720 & V & 16.3 & M07 & 10.4 & $<$38.5 &               \\
                   VCC 944 & V & 16.0 & M07 & 10.4 & $<$38.5 &               \\
                   VCC 369 & V & 15.8 & M07 & 10.4 &  39.2   &               \\
                  NGC 6017 & F & 29.5 &  HL & 10.3 &  39.3   &               \\
                  VCC 2000 & V & 15.0 & M07 & 10.3 &  38.6   &               \\
                  NGC 1172 & F & 22.0 &  HL & 10.3 &  38.5   &  NSC + X-ray  \\
                   VCC 654 & V & 14.7 & M07 & 10.3 & $<$38.4 &               \\
                  NGC 3377 & F & 10.4 &  HL & 10.3 &  38.6   &               \\
                   VCC 828 & V & 17.9 & M07 & 10.3 & $<$38.7 &               \\
                   VCC 778 & V & 17.8 & M07 & 10.3 &  38.6   &               \\
                   VCC 784 & V & 15.8 & M07 & 10.3 &  38.6   &  NSC + X-ray  \\
                  VCC 1250 & V & 17.6 & M07 & 10.3 &  38.8   &  NSC + X-ray  \\
                  VCC 1242 & V & 15.6 & M07 & 10.2 & $<$38.5 &               \\
                   VCC 355 & V & 15.4 & M07 & 10.2 &  38.7   &               \\
                  NGC 4742 & F & 15.3 &  HL & 10.2 &  39.2   &               \\
                  NGC 2778 & F & 22.7 &  HL & 10.2 &  38.7   & No HST; NSC?  \\
                  NGC 3457 & F & 20.5 &  HL & 10.2 &  38.8   & No HST; NSC?  \\
                  VCC 1630 & V & 16.1 & M07 & 10.2 & $<$38.3 &               \\
                  VCC 1327 & V & 18.3 & M07 & 10.2 &  38.8   &               \\
                  VCC 1913 & V & 17.4 & M07 & 10.2 & $<$38.5 &               \\
                  VCC 1619 & V & 15.5 & M07 & 10.2 &  38.6   &  NSC + X-ray  \\
                  VCC 1283 & V & 17.4 & M07 & 10.2 &  38.6   &  NSC + X-ray  \\
                  VCC 1303 & V & 16.8 & M07 & 10.1 & $<$38.3 &               \\
                   IC 1729 & F & 19.5 &   z & 10.1 &  39.0   &               \\
                   VCC 698 & V & 18.7 & M07 & 10.1 & $<$38.5 &               \\
                  VCC 1537 & V & 15.8 & M07 & 10.1 &  38.5   &               \\
                  NGC 4283 & F & 15.6 &  HL & 10.1 &  38.8   & No HST; NSC?  \\
                  VCC 1321 & V & 15.4 & M07 & 10.1 & $<$38.3 &               \\
               ESO 576-076 & F & 23.6 &   z & 10.1 & $<$38.4 &               \\
                 UGC 07767 & F & 27.5 &  HL & 10.0 &  38.7   &               \\
                  NGC 3641 & F & 26.4 &  HL & 10.0 &  38.8   & No HST; NSC?  \\
                  VCC 1146 & V & 16.4 & M07 & 10.0 & $<$38.3 &               \\
                  NGC 3522 & F & 25.5 &  HL &  9.9 &  38.8   & No HST; NSC?  \\
                  VCC 1475 & V & 16.6 & M07 &  9.9 & $<$38.4 &               \\
                  VCC 1125 & V & 16.4 & Vir &  9.9 & $<$38.5 &               \\
                  VCC 1261 & V & 18.1 & M07 &  9.9 & $<$38.5 &               \\
                  VCC 1178 & V & 15.8 & M07 &  9.9 &  38.6   &               \\
                  NGC 3073 & F & 33.4 &  HL &  9.8 & $<$38.9 &               \\
                  NGC 4121 & F & 24.8 &   z &  9.8 &  38.1   &               \\
                  NGC 1331 & F & 22.9 &  HL &  9.8 &  38.3   &  NSC + X-ray  \\
                 UGC0 5955 & F & 22.4 &   z &  9.7 & $<$38.4 &               \\
                     VCC 9 & V & 17.1 & M07 &  9.7 & $<$38.3 &               \\
                   VCC 571 & V & 23.8 & M07 &  9.7 & $<$38.8 &               \\
                  VCC 1297 & V & 16.3 & M07 &  9.7 &  38.4   &               \\
                   VCC 437 & V & 17.1 & M07 &  9.6 & $<$38.4 &               \\
                  VCC 1087 & V & 16.7 & M07 &  9.6 & $<$38.3 &               \\
                  VCC 2048 & V & 16.4 & Vir &  9.6 & $<$38.3 &               \\
                  NGC 1370 & F & 13.2 &   z &  9.6 &  38.7   & No HST; NSC?  \\
                  NGC 2970 & F & 25.9 &   z &  9.6 &  38.7   &  NSC + X-ray  \\
                  VCC 1422 & V & 15.3 & M07 &  9.5 & $<$38.2 &               \\
                 NGC 1097A & F & 16.7 &   z &  9.5 & $<$38.1 &               \\
                PGC 056821 & F & 27.0 &   z &  9.5 &  38.6   &               \\
                   VCC 856 & V & 16.8 & M07 &  9.5 & $<$38.3 &               \\
                  VCC 1695 & V & 16.5 & M07 &  9.5 & $<$38.3 &               \\
                  VCC 1431 & V & 16.1 & M07 &  9.5 & $<$38.6 &               \\
                  VCC 1861 & V & 16.1 & M07 &  9.5 & $<$38.3 &               \\
                  VCC 1192 & V & 16.1 &  HL &  9.5 & $<$38.7 &               \\
                  VCC 1910 & V & 16.1 & M07 &  9.5 & $<$38.3 &               \\
                  VCC 1871 & V & 15.5 & M07 &  9.4 & $<$38.2 &               \\
                  VCC 2019 & V & 17.1 & M07 &  9.4 & $<$38.4 &               \\
                  VCC 1355 & V & 16.9 & M07 &  9.4 &  38.6   &  NSC + X-ray  \\
                   VCC 140 & V & 16.4 & M07 &  9.4 & $<$38.3 &               \\
                   VCC 751 & V & 15.8 & M07 &  9.4 & $<$38.3 &               \\
                   VCC 543 & V & 15.7 & M07 &  9.4 & $<$38.2 &               \\
                  VCC 1512 & V & 18.4 & M07 &  9.3 & $<$38.3 &               \\
                  NGC 4308 & F & 12.0 &   z &  9.3 & $<$38.0 &               \\
                  VCC 1833 & V & 16.2 & M07 &  9.3 & $<$38.2 &               \\
                  VCC 1528 & V & 16.3 & M07 &  9.3 & $<$38.3 &               \\
                   VCC 200 & V & 18.2 & M07 &  9.3 & $<$38.5 &               \\
               PGC 3119319 & F & 23.6 &   z &  9.3 & $<$38.1 &               \\
                PGC 042748 & F & 15.5 &   z &  9.3 & $<$38.3 &               \\
                  VCC 1545 & V & 16.8 & M07 &  9.2 & $<$38.3 &               \\
                  VCC 1075 & V & 16.1 & M07 &  9.2 & $<$38.3 &               \\
                   VCC 538 & V & 22.9 & M07 &  9.2 & $<$38.7 &               \\
                  VCC 1440 & V & 16.0 & M07 &  9.2 & $<$38.2 &               \\
                  NGC 0855 & F &  9.6 &  HL &  9.2 &  38.6   &  Starforming  \\
                   IC 0225 & F & 21.9 &   z &  9.1 & $<$38.3 &               \\
                  VCC 1185 & V & 16.9 & M07 &  9.1 & $<$38.4 &               \\
                  VCC 1407 & V & 16.8 & M07 &  9.1 & $<$38.4 &               \\
                  NGC 7077 & F & 17.8 &   z &  9.1 & $<$38.3 &               \\
                  VCC 1627 & V & 15.6 & M07 &  9.1 & $<$38.3 &               \\
                  VCC 1993 & V & 16.5 & M07 &  9.0 & $<$38.3 &               \\
                  VCC 1488 & V & 16.4 & Vir &  9.0 & $<$38.3 &               \\
                    VCC 21 & V & 16.4 & Vir &  9.0 & $<$38.3 &               \\
                  VCC 1779 & V & 16.4 & Vir &  9.0 & $<$38.3 &               \\
                  VCC 1049 & V & 16.0 & M07 &  9.0 & $<$38.2 &               \\
                PGC 132768 & F & 20.1 &   z &  9.0 & $<$38.3 &               \\
                  VCC 1199 & V & 16.0 &  HL &  9.0 & $<$38.3 &               \\
                  VCC 1895 & V & 15.8 & M07 &  9.0 & $<$38.2 &               \\
                  VCC 2050 & V & 15.8 & M07 &  9.0 & $<$38.2 &               \\
                   VCC 230 & V & 17.8 & M07 &  9.0 & $<$38.7 &               \\
                  VCC 1661 & V & 15.8 & M07 &  9.0 & $<$38.2 &               \\
                  VCC 1743 & V & 17.6 & M07 &  9.0 & $<$38.3 &               \\
                  NGC 5099 & F & 19.0 &   z &  8.9 & $<$38.3 &               \\
                  VCC 1539 & V & 16.9 & M07 &  8.9 & $<$38.3 &               \\
               PGC 1210284 & F & 26.6 &   z &  8.8 & $<$38.4 &               \\
                    VCC 33 & V & 15.1 & M07 &  8.8 & $<$38.2 &               \\
                  VCC 1886 & V & 16.4 & Vir &  8.8 & $<$38.3 &               \\
                  VCC 1948 & V & 16.4 & Vir &  8.8 & $<$38.3 &               \\
                  VCC 1499 & V & 16.4 & Vir &  8.8 &  38.4   &  Starforming  \\
               PGC 1209872 & F & 26.6 &   z &  8.8 & $<$38.3 &               \\
                  VCC 1826 & V & 16.2 & M07 &  8.8 & $<$38.3 &               \\
                PGC 740586 & F & 19.0 &   z &  8.7 & $<$38.2 &               \\
                PGC 028305 & F & 23.0 &   z &  8.7 & $<$38.3 &               \\
                  VCC 1489 & V & 16.5 &  HL &  8.7 & $<$38.3 &               \\
               PGC 1242097 & F & 27.7 &   z &  8.7 & $<$38.4 &               \\
               ESO 540-014 & F & 22.4 &   z &  8.7 &  40.1   &  Starforming  \\
                PGC 042173 & F & 23.0 &   z &  8.6 & $<$38.3 &               \\
                PGC 064718 & F &  9.7 &   z &  8.6 & $<$38.2 &               \\
                PGC 042737 & F & 26.6 &   z &  8.6 & $<$38.4 &               \\
               PGC 1216386 & F & 26.6 &   z &  8.5 & $<$38.3 &               \\
               PGC 1230503 & F & 27.7 &   z &  8.5 & $<$38.4 &               \\
                PGC 030133 & F & 17.2 &   z &  8.5 & $<$38.3 &               \\
       6dF J2049400-324154 & F & 24.2 &   z &  8.4 & $<$38.4 &               \\
               PGC 1202458 & F & 25.4 &   z &  8.4 & $<$38.3 &               \\
     SDSS J145828.64+01323 & F & 23.0 &   z &  8.3 & $<$38.3 &               \\
     SDSS J150812.35+01295 & F & 23.6 &   z &  8.3 & $<$38.3 &               \\
                PGC 042724 & F & 10.9 &   z &  8.2 & $<$38.3 &               \\
     SDSS J150907.83+00432 & F & 25.4 &   z &  8.1 & $<$38.4 &               \\
               PGC 1179083 & F & 25.4 &   z &  8.1 & $<$38.4 &               \\
                PGC 042596 & F & 12.6 &   z &  8.0 & $<$38.3 &               \\
               PGC 3097911 & F & 19.0 &   z &  8.0 & $<$38.3 &               \\
                PGC1 35659 & F & 15.5 &   z &  8.0 & $<$38.5 &               \\
               PGC 1206166 & F & 26.6 &   z &  8.0 &  38.7   & No HST; NSC?  \\
     SDSS J150233.03+01560 & F & 25.4 &   z &  8.0 & $<$38.3 &               \\
     SDSS J150100.85+01004 & F & 26.6 &   z &  7.9 & $<$38.3 &               \\
               PGC 1223766 & F & 24.2 &   z &  7.9 & $<$38.3 &               \\
                PGC 135818 & F & 14.9 &   z &  7.9 & $<$38.3 &               \\
                PGC 135829 & F & 20.7 &   z &  7.9 & $<$38.3 &               \\
               PGC 1217593 & F & 17.2 &   z &  7.9 & $<$38.3 &               \\
     SDSS J145944.77+02075 & F & 22.4 &   z &  7.9 & $<$38.3 &               \\
                PGC 042454 & F & 12.6 &   z &  7.9 & $<$38.3 &               \\
                PGC 085239 & F & 20.7 &   z &  7.8 & $<$38.3 &               \\
     SDSS J150033.02+02134 & F & 20.1 &   z &  7.8 & $<$38.3 &               \\
               PGC 1192611 & F & 23.6 &   z &  7.8 & $<$38.4 &               \\
                PGC 043421 & F & 16.7 &   z &  7.7 & $<$38.3 &               \\
\enddata                   
\vspace{-0.3cm}            

\tablecomments{Column 1: Object name from HyperLeda, or VCC for Virgo
  galaxies; Column 2: V = Virgo, F = Field; Column 3: Adopted distance
  in Mpc; Column 4: Distance method, with M07 = from Mei et
  al.~(2007), HL = non-redshift distance in HyperLeda, z = calculated
  from redshift, and Vir = assumed 16.4 Mpc for Virgo; Column 5:
  Stellar mass calculated as in G10 and M12a for these distances;
  Column 6: X-ray luminosity calculated as in G10 and M12a for these
  distances; Column 7: Starforming = clumps of star formation present,
  removed from clean sample; NSC + X-ray = dual nuclear star cluster
  and nuclear X-ray source; No HST; NST? = lacks high-resolution ACS
  {\it HST\/} coverage but might be a candidate for LMXB contamination
  of the nuclear X-ray emission if a NSC is present. See text for
  details of construction of the safe sample.}

\end{deluxetable*}

\end{document}